\documentclass[12pt,authoryear]{article}

\usepackage{amsmath,amsfonts,amssymb}

\usepackage{lineno}
\usepackage{pgf}
\usepackage{tikz}
\usepackage{subfigure}
\usepackage{algorithm}
\usepackage{algorithmic}
\usepackage[colorlinks=true]{hyperref}
\usepackage{color}
\begin{document}




\title{About fluid forces computation for Volume Penalization coupled with Lattice Boltzmann method (VP-LBM)}


\author{Erwan Liberge, Claudine B\'eghein \\
LaSIE UMR 7356 CNRS, Universit\'e de La Rochelle,\\  Avenue Michel Cr\'epeau, 17042 La Rochelle Cedex \\ France}
\date{eliberge@univ-lr.fr}
\maketitle
\begin{abstract}
In this paper, two different approaches fluid forces computation approaches are compared in the frame of Volume Penalization - Lattice Boltzmann Method (VP-LBM). The first method, the momentum exchange method, uses the variation of the distribution functions near the fluid solid interface, while the second one, the stress integration method, allows direct integration of fluid forces onto this interface. Applied to the VP-LBM, which consists in penalizing the solid in the LBM, these two methods lead to significantly different results. The tests are performed to study on one hand the lift and drag coefficients of a Naca 0012 airfoil at different angles of attack, at Reynolds number 1000, and on the other hand, the particle sedimentation under gravity in a channel. 

\end{abstract}

\begin{paragraph}{Keyword}
Lattice Boltzmann Method, Fluid Structure Interaction, Volume Penalization, Momentum Exchange Method, Stress Integration Method
\end{paragraph}




\section{Introduction}
\label{sec:1}
Computational modelling of fluid-structure interaction (FSI) has remained a challenging research area over the past few decades. Many efficient methodologies and algorithms to model FSI have evolved in the recent past.  A classical approach consists in coupling a fluid solver for the Navier-Stokes equations  with a structure solver, the fluid solver being obtained by a classical discretization method, such as the finite element or the finite volume method. We propose in this paper to use the Lattice Boltzmann Method (LBM) as a fluid solver for FSI simulation. 

The LBM  has been successfully developed for computational fluid mechanics since the 90's \cite{Benzi1992145} and appears to be an alternative computational method. Based on the Boltzmann equation, the LBM considers the transport of the probability to find a particle according to time, space and velocity; the Boltzmann equation being solved according to space, velocity and time. The macroscopic variables are obtained using moments of the distribution functions. The power of the LBM resides in its programming simplicity and the short computational time if the algorithm is solved using Graphic Processor Units (GPU) \cite{Fan2004297}. LBM approaches for solving flows around moving bodies can be classified in two families. 

The first one concerns the Bounce-Back methods and their derivatives. The Bounce-Back methods consist in considering that a wall rejects the particle, and, for a moving boundary, in changing locally the macroscopic velocity. For moving bodies, this family can be decomposed in $4$ groups as suggested by Kr\"uger et al. \cite{krugerbook}. In the first group of methods, the boundary is approximated in a staircase manner \cite{Ladd20011191}. This method can lead to errors in case of complex geometries, and for moving boundaries, it needs an expensive step for updating the fluid site and a refilling algorithm on nodes which become fluid. The second group deals with methods which use interpolation to impose the exact wall velocity \cite{Yu20032003,Bouzidi20013452}. The results obtained with such methods are more accurate but have one drawback due to the interpolation: the mass is not conserved. The other drawback is the use of a fulfill algorithm to compute quantities on solid nodes which become fluid after the boundary movement. The following group focuses on methods called Partially Saturated Bounce-Back (PSBB) in Kruger \textit{et al.} \cite{krugerbook}. The principle is that a lattice node can be a mixed fluid/solid node. The method, originally proposed by Noble and Torczynski \cite{Noble19981189} consists in changing the collision operator by introducing a volume fraction of the solid. Finally, the collision operator is a mixing between the classical collision operator and the Bounce-Back method. The major drawback of this method is the difficulty to compute the volume fraction of solid for each lattice node. This restrains the domain of application of this method to stationary bodies. Kr\"uger et al. \cite{krugerbook} propose a last group of methods based on the extrapolation of the distribution functions for the fluid nodes located near the boundary. 

The second family is the Immersed-Boundary (IB) methods for LBM \cite{Feng2004602} which consists in modelling the effect of the boundary by adding nodal forces in the vicinity of the boundary, in the fluid flow solver.  The principal drawback of the IB-LBM is that the nodal forces use a penalization factor, and the hydrodynamic forces and torques depending on this factor for rigid bodies. The Direct forcing scheme \cite{Dupuis20084486} cancels this drawback, but it requires to solve the Boltzmann equation twice per time step. Wang et al. \cite{Wang2015440} propose another approach using a Lattice Boltzmann Flux Solver (LBFS), whose formulation is not efficient for GPU implementation.

In previous papers \cite{Benamour2015299, Benamour2017481,BENAMOUR2020101050}, we proposed to couple the Volume Penalization (VP)  method \cite{Angot1999497} and the LBM (VP-LBM). The Volume Penalization method consists in extending the Navier-Stokes equations to the whole domain (fluid and solid) and in adding a volume penalization term to take account the structure. The approach can be seen as a mix between the Partially Saturated Bounce-Back (PSBB) and the Immersed Boundary (IB) methods. However, the Volume Penalization method does not require the expensive computation of the solid fraction near the solid interface as in PSBB methods, and the difference compared to the IB methods is that the VP method uses a volume force, instead of local forces on Lagrangian markers. The ability of penalty methods for fluid structure interaction problems has been demonstrated by Destuynder \textit{et al.} \cite{DESTUYNDER20211}. In a previous works Benamour \textit{et al.} \cite{Benamour2015299, Benamour2017481} showed that the VP-LBM gives good results for fixed bodies. In \cite{BENAMOUR2020101050}, the method has been successfully tested for moving boundaries and a real case of fluid structure interaction (FSI). In this previous work, the Momentum Exchange (ME) method was used to compute fluid forces, and, although results were validated, spurious oscillations could have been observed for the FSI case on lift and drag coefficients. We propose in this paper, to compare  ME and the other well-known method to compute forces in LBM, the Stress-Integration (SI) method on new cases, the NACA 0012 airfoil with angles of attack from $0^{\circ}$ to $28^{\circ}$ at Reynolds number $1000$, and the particle sedimentation under gravity in a channel.    

The theoretical background is presented in the following section. The present part deals with the Lattice Boltzmann Method and more particularly the Two Relaxation Time (TRT) approach, the Volume Penalization and how the combination of these two methods. Then, the Momentum Exchange (ME) and Stress Integration (SI) methods are introduced. The last section presents the applications computed on a GPU device. For the first case tested, the lift and drag coefficients of a Naca 0012 airfoil at different angles of attack, at Reynolds number 1000 obtained with ME and SI are compared. The second example deals with particle sedimentation under gravity in a channel.

\section{Governing equations}
\label{sec:2}
In this section, the numerical models are exposed. The following notations are used : $\rho$ and $\mathbf{u}$ are the macroscopic density and velocity, and bold characters denote vectors.
\subsection{Volume penalization}
Let us consider a fluid domain $\Omega_f$, a solid domain $\Omega_s$, $\Gamma$ the fluid-solid interface, and let us note $\Omega = \Omega_f \cup \Omega_s \cup \Gamma$. The Volume Penalization (VP) method consists in extending the Navier-Stokes equations to the whole domain $\Omega$, and considering the solid domain as a porous medium with a very small permeability. The method was introduced by Angot et al. \cite{Angot1999497} and already applied to macroscopic equations for moving bodies  \cite{Kadoch20124365}. The small permeability of the solid domain is  modelled using a penalization coefficient, hence the desired boundary conditions at the fluid-solid interface are naturally imposed. With this method, the incompressible Navier-Stokes equations are written as follows :
\begin{equation}
\begin{array}{ll}
\displaystyle \nabla \cdot \mathbf{u}=0 \\
\displaystyle \frac{\partial \mathbf{u}}{\partial t}+\mathbf{u}\cdot \nabla \mathbf{u} = \displaystyle- \frac{1}{\rho}\nabla p + \nu \nabla^2 \mathbf{u} -\dfrac{\chi_{\Omega_s}}{\eta} \left(\mathbf{u} - \mathbf{u_s}\right)  \\
\end{array}
\end{equation} 
where  
\begin{equation}
\chi_{\Omega_s}\left(\mathbf{x},t\right)=\left\lbrace \begin{array}{ll} 1 \;\; \textrm{if} \;\; \mathbf{x} \in \Omega_s\left( t \right) \\
0 \;\; \textrm{otherwise}
\end{array}
 \right. ;  \hspace{1cm}\eta \ll 1\;\; \textrm{penalization factor}
 \label{eq:fcar}\end{equation} 
$\mathbf{u}$ denotes the velocity field, $p$ is the pressure field, $\rho$ and $\nu$ are the density and the viscosity of the fluid. $\mathbf{F}=\dfrac{\chi_{\Omega_s}}{\eta} \left(\mathbf{u} - \mathbf{u_s}\right)  $ is the penalization term, and $ \mathbf{u_s}$ is the velocity field in the solid domain. 
\subsection{Lattice Boltzmann method}
Based on the Boltzmann equation (equation (\ref{eqB}))  proposed in the context of the Kinetic Gaz Theory by L. Boltzmann in 1870, the Lattice Boltzmann Method has been successfully used  to model fluid flow since the 90's. 
\begin{equation}
\label{eqB}
\displaystyle \dfrac{\partial f }{\partial t } + \mathbf{c}\cdot \nabla_x f= \Omega\left(f \right) 
\end{equation}
This equation models the transport of  $f\left(\mathbf{x},t,\mathbf{c}\right)$, a probability density function of particles with the velocity $\mathbf{c}$ at location $\mathbf{x}$ and time $t$. $\Omega\left(f \right) $ is the collision operator. The link between the Boltzmann equation and the Navier-Stokes equations is well-known since the Chapmann-Enskog expansion proposed in 1915. \\ 
The Lattice Boltzmann method considers the discretization of equation (\ref{eqB}) according to space and velocity and leads to the following discretized equations :
\begin{equation}
\label{eqBd}
\displaystyle f_{\alpha} \left(\mathbf{x}+\mathbf{c_\alpha}\triangle t, t +\triangle t \right) -f_{\alpha} \left(\mathbf{x}, t  \right) = \Omega_\alpha \left(f \right) +\triangle t F_\alpha 
\end{equation}
where $ \displaystyle  f_{\alpha} \left(\mathbf{x}, t  \right) = \displaystyle f \left(\mathbf{x},\mathbf{c_{\alpha}}, t  \right) $, $ \displaystyle F_\alpha $ is a forcing term related to the discrete velocity $\mathbf{c_\alpha}$ \cite{guo_discrete_2002} . 
 
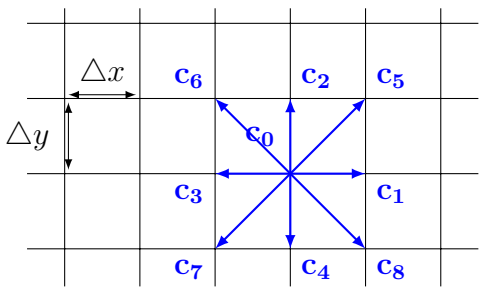
\begin{figure}[!htbp]
\center 
\begin{tikzpicture}
\draw [latex-latex] (0.05,2.05) -- (0.95,2.05);
\node [above] at (0.5,2) { $\triangle x$};
\draw [latex-latex] (0.05,1.05) -- (0.05,1.95);
\node   at (-0.5,1.5) { $\triangle y$};
\node  [blue] at (2.6,1.5) {$\mathbf{c_0}$};
\node  [blue,below right] at (4,1) {$\mathbf{c_1}$};
\node  [blue,above right] at (3,2) { $\mathbf{c_2}$};
\node  [blue,below left] at (2,1) { $\mathbf{c_3}$};
\node  [blue,below right] at (3,0) { $\mathbf{c_4}$};
\node  [blue,above right] at (4,2) {$\mathbf{c_5}$};
\node  [blue,above left] at (2,2) { $\mathbf{c_6}$};
\node  [blue,below left] at (2,0) { $\mathbf{c_7}$};
\node  [blue,below right] at (4,0) { $\mathbf{c_8}$};
\draw (-0.5,-0.5) grid (5.5,3.2);
\draw [blue,thick,-latex] (3,1) -- (4,1);
\draw [blue,thick,-latex] (3,1) -- (4,2);
\draw [blue,thick,-latex] (3,1) -- (3,2);
\draw [blue,thick,-latex] (3,1) -- (2,2);
\draw [blue,thick,-latex] (3,1) -- (2,1);
\draw [blue,thick,-latex] (3,1) -- (2,0);
\draw [blue,thick,-latex] (3,1) -- (3,0);
\draw [blue,thick,-latex] (3,1) -- (4,0);
\end{tikzpicture}
\caption{\label{d2q9} Discrete velocities of the D2Q9 model}
\end{figure}
\begin{equation}
\mathbf{c_\alpha} = \left\lbrace
\begin{array}{lll}
\left(0,0 \right) & \alpha =0\\
\left( \textrm{cos} \left(   \left(  \alpha -1 \right)  \dfrac{\pi}{2} \right) ,  \textrm{sin} \left(   \left(  \alpha -1 \right)  \dfrac{\pi}{2} \right) \right)c & \alpha =1,2,3,4 \\
\left( \textrm{cos} \left(   \left( 2 \alpha -9 \right)  \dfrac{\pi}{4} \right) ,  \textrm{sin} \left(   \left(2  \alpha -9 \right)  \dfrac{\pi}{4} \right) \right)\sqrt{2}c& \alpha =5,6,7,8
\end{array}
 \right.
\end{equation}
Where $c=\dfrac{\triangle x }{\triangle t}$. Usually $\triangle x=\triangle y= \triangle t=1$ are chosen.

 The first model proposed by Bhatnagar et al. \cite{Bhatnagar1954511} is the BGK model which is based on a linear collision operator with a single relaxation time :
\begin{equation}
\Omega_\alpha \left(f\right)= -\dfrac{1}{\tau} \left(f_\alpha \left( \mathbf{x},t\right) -f ^{\textrm{eq}}_\alpha \left( \mathbf{x},t\right) \right)
\end{equation}
where $f ^{\textrm{eq}}$ is the equilibrium function, 
\begin{equation}
\displaystyle f^{\textrm{eq}}_\alpha  = \displaystyle \omega_\alpha \rho \left( 1 + \dfrac{\mathbf{c_\alpha} \cdot \mathbf{u}}{c^2_s} +  \dfrac{ \mathbf{u}  \mathbf{u}  :  \left( \mathbf{c_\alpha} \mathbf{c_\alpha} - c^2_s I \right) }{2c^4_s}  \right),
\end{equation}
$\omega_\alpha=\left\lbrace 4/9,1/9,1/9,1/9,1/9,1/36,1/36,1/36,1/36\right\rbrace$, $c_s=\dfrac{c}{ \sqrt{3}}$ and $\tau$ is the non dimensional relaxation time which is linked to the fluid viscosity as follows.
\begin{equation}\label{tauvisco}
\nu =c^2_s\triangle t \left( \tau -\dfrac{1}{2}\right)
\end{equation} 

In order to increase the stability, approaches using multiple relaxation times have been proposed \cite{dh92,krugerbook}. In this work, the Two Relaxation Times (TRT) method is used. 

We note $\mathbf{c_\alpha}$ the discrete velocity according the direction $\alpha$ and $\mathbf{c_{\bar{\alpha}}}=-\mathbf{c_\alpha}$ the discrete velocity in the opposite direction $\bar{\alpha}$.

Then, the TRT method leads to introduce positive and negative modified distribution functions :
\begin{equation}
 f_\alpha^{ +}=\frac{f_\alpha+f_{\bar{\alpha}}}{2}~,~f_\alpha^{ -}=\frac{f_\alpha-f_{\bar{\alpha}}}{2}.
\end{equation}
In the same way, are defined $f_\alpha^{eq ~+}$ and $f_\alpha^{eq ~-}$.

This leads to the following discretised scheme :
\begin{multline} 
f_\alpha \left( \mathbf{x}+\mathbf{c}_\alpha \Delta t, t+\Delta t \right)-f_\alpha \left(\mathbf{x},t\right)=-\frac{\Delta t}{\tau^{+}}\left(f_\alpha^{+}\left(\mathbf{x},t\right)-f_\alpha^{eq ~+}\left(\mathbf{x},t\right)\right) 
\\ -\frac{\Delta t}{\tau^{-}}\left(f_\alpha^{-}\left(\mathbf{x},t\right)-f_\alpha^{eq ~-}\left(\mathbf{x},t\right)\right) + \left(1-\frac{\Delta t}{2 \tau^{+}}\right) F_\alpha ,
\label{TRT}
\end{multline}
where $\tau^+$ is the relaxation time linked with the non dimensional viscosity $\nu$ according to:
\begin{equation} 
\nu=c^2_s\triangle t \left(\tau^{+}-\frac{1}{2}\right) .
\end{equation}
The relaxation time $\tau^{-}$ is obtained as follows:
\begin{equation}
\tau^{-}=\frac{\Delta t~\Lambda}{\tau^{+}-\frac{1}{2}}+\frac{1}{2},
\end{equation}
In this work, we choose $\Lambda=\frac{1}{6}$ , due to the best stability we obtained with this value.

Finally, the macroscopic quantities are computed according to the following expressions :
\begin{equation}\label{updatestep}
\rho=\sum_\alpha f_\alpha \hspace{1cm} \rho \boldsymbol{u}=\sum_\alpha \mathbf{c_\alpha} f_\alpha +\dfrac{\triangle t}{2} \rho \mathbf{F}  
\end{equation}
In the present approach, the volume penalization term is added :
\begin{equation}\label{updatestep2}
 \rho \mathbf{u}=\sum_\alpha \mathbf{c_\alpha} f_\alpha -\dfrac{\triangle t}{2} \rho \dfrac{\chi_{\Omega_s}}{\eta} \left(\mathbf{u} - \mathbf{u_s}\right)  
\end{equation}
To avoid instabilities, the term including $\mathbf{u}$ in the penalization force is moved to the left hand side of equation (\ref{updatestep2}) 
\begin{equation}\label{updatestep3}
 \rho \left(1 + \dfrac{\triangle t}{2} \dfrac{\chi_{\Omega_s}}{\eta}  \right) \mathbf{u}=\sum_\alpha \mathbf{c_\alpha} f_\alpha +\dfrac{\triangle t}{2} \rho \dfrac{\chi_{\Omega_s}}{\eta}  \boldsymbol{u_s}  
\end{equation}

 This leads to the modified update step to compute the macroscopic velocity field :
\begin{equation}\label{updatestepm}
 \displaystyle \mathbf{u}= \displaystyle  \dfrac{ \displaystyle \sum_\alpha \mathbf{c_\alpha} f_\alpha   +   \dfrac{\triangle t}{2}  \dfrac{\chi_{\Omega_s}}{\eta}  \rho \mathbf{u_s}  }{\rho +    \dfrac{\triangle t}{2}  \dfrac{\chi_{\Omega_s}}{\eta}  \rho } 
\end{equation}
In the fluid domain, where $ \chi_{\Omega_s} = 0$ the classical LBM equation is obtained whereas in the solid domain, where $ \chi_{\Omega_s} = 1$, equation (\ref{updatestepm}) forces the velocity field to approach $ \mathbf{u_s}$.

\subsection{Fluid forces computation}

Angot \textit{et al.} \cite{Angot1999497} proposed in a context of an integral formulation of the volume penalization problem to compute the fluid forces with the following formula :
\begin{equation} \label{Angotforce}
\boldsymbol{\mathcal{F}_f}=\underset{\eta \longrightarrow\infty}{\textrm{lim}} \int_{\Omega_s} \mathbf{u} -\mathbf{u_s} d\Omega
\end{equation}
The formula (\ref{Angotforce}) works with finite element or finite volume methods, but fails on our computational tests. We present in the following the two classical method used in LBM to compute fluid forces.
\subsubsection{Momentum Exchange Method (MEM) \label{memsection}}
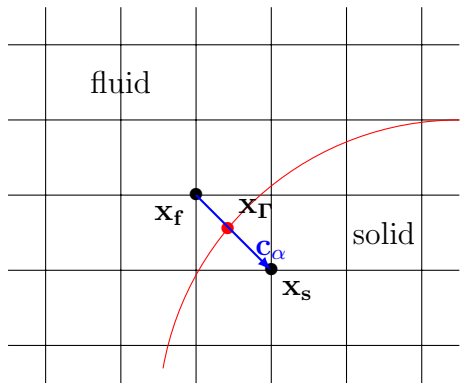
\begin{figure}[!htpb]
  \centering
 \begin{tikzpicture}
  \draw (-0.5,-0.5) grid (5.5,4.5);
  \draw  [red] (5.5,3) arc [radius=4, start angle=90, end angle= 170];
  \node [below right] at (3,1) {$\mathbf{x_s}$};
  \node at (3,1) {$\bullet$};
\node [below left] at (2,2) {$\mathbf{x_f}$};
   \node at (2,2) {$\bullet$};
   \node [above right] at (2.42,1.55) {$\mathbf{\displaystyle x_\Gamma}$};
      \node [red] at (2.42,1.55) {$\bullet$};
  \draw [blue,thick,-latex] (2,2) -- (3,1);
  \node [blue,above] at (3,1) {$\mathbf{c_{\alpha}}$};
  \node at (1,3.5) {fluid};
  \node at (4.5,1.5) {solid};
\end{tikzpicture}
\caption{Curved interface on a square lattice : example of a fluid boundary node $\mathbf{x_f}$, its image in the solid domain $\mathbf{x_s}$, and the intersection point $ \mathbf{\displaystyle x_\Gamma}$ located on the interface}
  \label{fig:BBdraft}
\end{figure}

The fluid forces are computed with the momentum exchange method (MEM) proposed by Wen at al.\cite{WEN2014161}.
We note $\mathbf{x_f}$ a boundary node in the fluid domain and $\mathbf{x_s}$ the image of this boundary node through the solid interface by a lattice velocity $\mathbf{c_\alpha}$, also called incoming velocity( cf. figure \ref{fig:BBdraft}). The intersection point between the fluid-solid interface and the link $\mathbf{x_f} -\mathbf{x_s} $ is $\displaystyle \mathbf{\displaystyle x_\Gamma}$, and the outgoing lattice velocity is denoted $\displaystyle \mathbf{ \displaystyle c_{ \overline{\alpha}}} = - \mathbf{c_\alpha}$.
 
 The local force at $\mathbf{\displaystyle x_\Gamma}$ is computed using the following expression :
\begin{equation}\label{MMEL}
\mathbf{F} \left(\mathbf{\displaystyle x_\Gamma} \right) = \left(\mathbf{c_\alpha} - \mathbf{u_\Gamma} \right) \tilde{f}_\alpha \left( \mathbf{x_f}\right) -\left(\mathbf{c_{ \overline{\alpha}}} - \mathbf{u_\Gamma} \right) \tilde{f}_{ \overline{\alpha}} \left( \mathbf{x_s}\right)  
\end{equation}
and the total fluid force acting on the solid domain is :
\begin{equation}
\boldsymbol{\mathcal{F}_f}= \displaystyle \sum \mathbf{F} \left(\mathbf{\displaystyle x_\Gamma} \right)
\end{equation}

The torque is obtained with
\begin{equation}\label{fluidtorque}
\boldsymbol{\mathcal{T}_f}=\displaystyle \sum \left(\mathbf{x_\Gamma} -  \mathbf{x_G} \right) \times  \mathbf{F} \left(\mathbf{\displaystyle x_\Gamma} \right),
\end{equation}
with $\mathbf{x_G}$ the coordinates of the gravity center of the body.

Giovacchini and Ortiz \cite{PhysRevE.92.063302} showed that the MEM does not depend of the way the boundary conditions at the solid domain are implemented.

\subsubsection{Stress Integration Method (SIM)}
This method is more intuitive in computational fluid dynamics, and consists in integrating the fluid stress tensor onto the structure structure:
\begin{equation}\label{defFf}
\boldsymbol{\mathcal{F}_f}= \displaystyle \int_{\partial \Omega_s}\sigma \cdot \mathbf{n} dS\; \textrm{and} \;\;  \boldsymbol{\mathcal{T}_f}=\displaystyle \int_{\partial \Omega_s} \mathbf{r} \times  \sigma \cdot \mathbf{n}dS
\end{equation}
with 
\begin{equation}\begin{array}{ll}
\sigma  &=  \displaystyle -p I_d+ \nu \left( \nabla \mathbf{u} + \left(\nabla \mathbf{u}\right)^T\right) \\
       & =   \displaystyle -\rho c^2_s I_d- \left(1-\dfrac{1}{2\tau}\right) \left(\sum_\alpha  \mathbf{c_\alpha} \otimes \mathbf{c_\alpha} \left(f_\alpha - f^{eq}_\alpha \right) \right)
\end{array}
\end{equation}
and $\mathbf{n}$ is the outward normal to the solid interface.

The $f_{\alpha}$ are extrapolated from the closest point in the relevant direction (close to $\mathbf{n}$) in the fluid domain to the integration points $\mathbf{x_i}$ located on the surface. Finally, the equation (\ref{defFf}) becomes :
\begin{equation}
\boldsymbol{\mathcal{F}_f}= \displaystyle \sum_i S_i \sigma\left(\mathbf{x_i}\right) \cdot \mathbf{n_i}
\end{equation} 
$S_i$ and $\mathbf{n_i}$ are the integration surface and the outward normal at integration point $\mathbf{x_i}$.

\section{Applications}
\label{sec:3}

All computations were run on a NVIDIA QUADRO P500 GPU card, using a CUDA implementation. 
A value of penalization factor $\eta = 10^{-6}$ was selected for all cases.

In the followings $\textrm{l.u.}$ refers to $\textrm{lattice length units}$ and $\textrm{t.s.}$ to $\textrm{lattice time units}$.


\subsection{NACA airfoil}
The first application is the study of the NACA 0012 airfoil with different angle attack values at Reynolds number $1000$. This case is well-documented in literature, and the different ME or SI results for VP-LBM are compared to those obtained by \cite{DIILIO2018200,Kurtulus2015,LIU20123427}. 

Liu \textit{et al.} \cite{LIU20123427} use the finite elements method combined with a fine mesh to give accurate numerical results. Kurtulus \cite{Kurtulus2015} proposes a very complete study, using finite volume method and a lot of data to compare. Di Illio \textit{et al.} \cite{DIILIO2018200} combine the standard LBM with an unstructured finite volume formulation in the so-called hybrid lattice Boltzmann method. They have used an overlap between a standard LBM approach on the whole domain and an unstructured body-fitted grid model where a finite-volume lattice Boltzmann formulation is applied. This approach has led to very accurate results close to the body. However, no information has been given on fluid forces calculation. It looks like a Stress Integration method because the macroscopic values have directly been taken from the body fitted mesh.

\begin{figure}[!htpb]
  \centering
  \begin{tikzpicture}
    \draw [thick] (0,0) -- (12,0);
 \draw [thick] (0,5) -- (12,5);
 \draw [thick] (4,2.5) .. controls (2.5,3.5) and (2.5,2.5) .. (4,2.5);
\draw [cyan,thick,-latex] (0,0) -- (0.5,0);
\draw [cyan,thick,-latex] (0,0.5) -- (0.5,0.5);
\draw [cyan,thick,-latex] (0,1) -- (0.5,1);
\draw [cyan, thick,-latex] (0,1.5) -- (0.5,1.5);
\draw [cyan, thick,-latex] (0,2) -- (0.5,2);
\draw [cyan, thick,-latex] (0,2.5) -- (0.5,2.5);
\draw [cyan,thick,-latex] (0,3) -- (0.5,3);
\draw [cyan,thick,-latex] (0,3.5) -- (0.5,3.5);
\draw [cyan,thick,-latex] (0,4) -- (0.5,4);
\draw [cyan,thick,-latex] (0,4.5) -- (0.5,4.5);
\draw [cyan,thick,-latex] (0,5) -- (0.5,5);
\draw [latex-latex] (4,2.5) --(2.9,2.9); 
\node [right] at (3.3,3.1) {$C$};
\draw [dashed] (4,2.5) --(2,3.2); 
\draw [dashed] (4,2.5) --(2,2.5); 
\node [left] at (2.4,2.8) {$\alpha$};
\draw[latex-latex] (2.5,3) arc[radius=0.3, start angle=110, end angle=220];
\draw [latex-latex] (0,1.75) -- (4,1.75);
\node [below] at (1.5,1.75) {$L_1$};
\draw [latex-latex] (4,1.75) -- (12,1.75);
\node [below] at (7.5,1.75) {$L_2$};
\draw [latex-latex] (12.5,0) -- (12.5,5);
\node [right] at (12.5,2.5) {$H$};
\node [left] at (0,2.5) {$U_0$};
\node [left] at (0,3) {inlet};
\node [left] at (12,3) {outlet};
\node [left] at (8,3) {symmetry};
\draw [-latex] (7,3.5) .. controls (7.1,4) and (6.75,4.45) .. (6,5);
\draw [-latex] (7,2.5) .. controls (7.1,2) and (7,1) .. (6,0);

\draw [latex-latex] (1.5,4.5)--(1.5,3.5) -- (2.5,3.5);
\node [above] at (2.5,3.5) {$\mathbf{x}$};
\node [right] at (1.5,4.5) {$\mathbf{y}$};
\end{tikzpicture}
\caption{Shematic of the computational domain around the NACA airfoil  \label{nacashem}}
\end{figure}
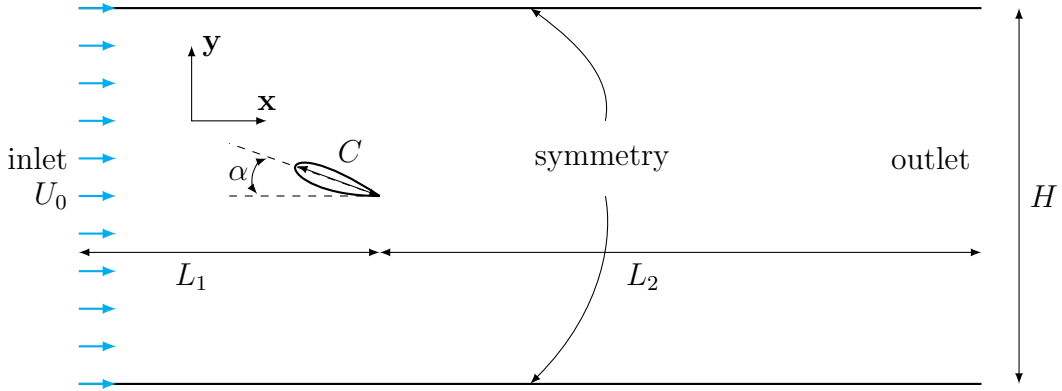

The figure \ref{nacashem} represents the computational domain. 
Let $C$ be the chord of the NACA 0012. The airfoil is placed at $4C$ from the inlet and $9C$ from the outlet. The height of the computational domain is $7C$, and the NACA is $3.5C$ from the bottom. 

A constant velocity profile has been imposed at the inlet using the classical half-way Bounce-Back method, and the outflow boundary condition at outlet has been modeled using the convective condition \cite{Yang2013160}. This condition makes it possible to reduce the distance between the airfoil and the extreme limit of the computational domain downstream the immersed body. Symmetry boundary conditions ( $\mathbf{u}\cdot \mathbf{n}=0$) have been imposed at the other boundaries. 

The computations have been carried out using the following parameters (in lattice units):
 $$ C=278, \;\; U_0= 0.0599,\;\; \tau = 0.55$$
 Note that $\tau$ is close to the stability limit for LBM, but this makes it possible to decrease $C$ and then the size of the computational problem and also the computational time. Di Illio \textit{et al.} \cite{DIILIO2018200} have used $512$ nodes in the chord, and a larger computational domain. However, our interest in VP-LBM solved in CUDA being the computational time, we try to obtain a good qualitative result without too expensive computing resources.  This is why this set of parameters have been used. 
 
For the Stress Integration method, 849 integrations points have been used. Note that the number of integration points has been chosen arbitrary. Increasing their number increases the accuracy of the computation, but not significantly in this case. 

\begin{figure}[!htbp]
\center 
\subfigure[\label{liftnaca} Lift coefficient]{\includegraphics[width=6.5cm]{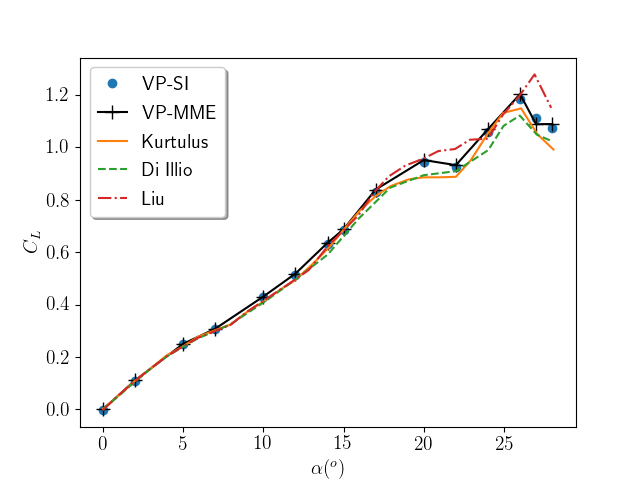}}
\subfigure[\label{dragnaca} Drag coefficient]{\includegraphics[width=6.5cm]{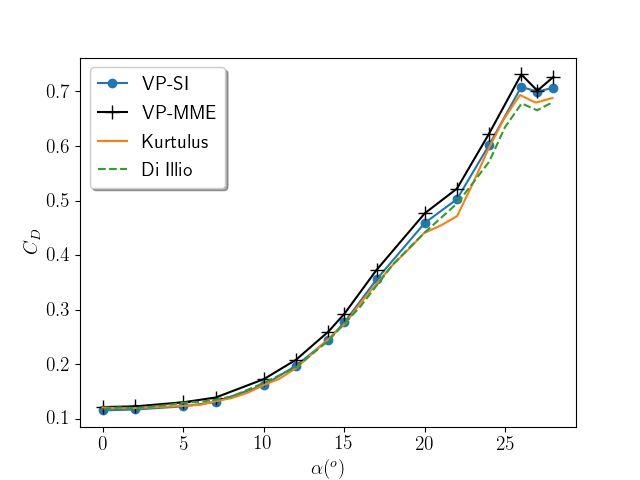}}
\caption{\label{NACAcdcls} Lift and Drag coefficient versus angle. The VP-LBM's results computed with Stress Integration and Momentum Exchange methods are compared with literature }
\end{figure}

The drag and lift coefficients are plotted in figure \ref{NACAcdcls}. The VP-LBM gives a good prediction of these values compared to  literature. 

\begin{figure}[!htbp]
\center 
\subfigure[\label{naca7} $\alpha=7^{\circ}$]{\includegraphics[width=6.5cm]{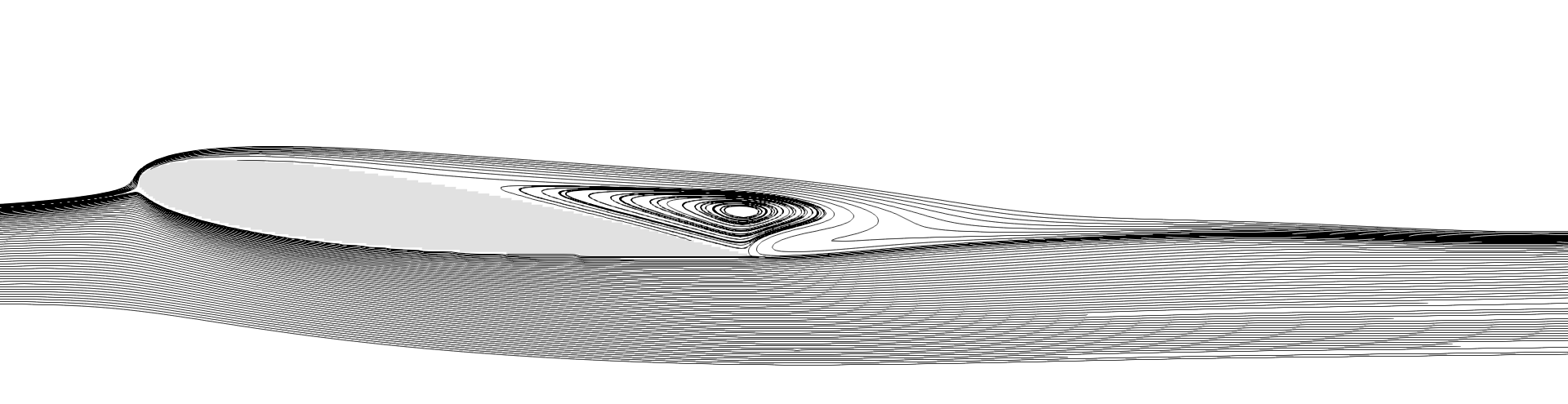}}
\subfigure[\label{naca10} $\alpha=10^{\circ}$]{\includegraphics[width=6.5cm]{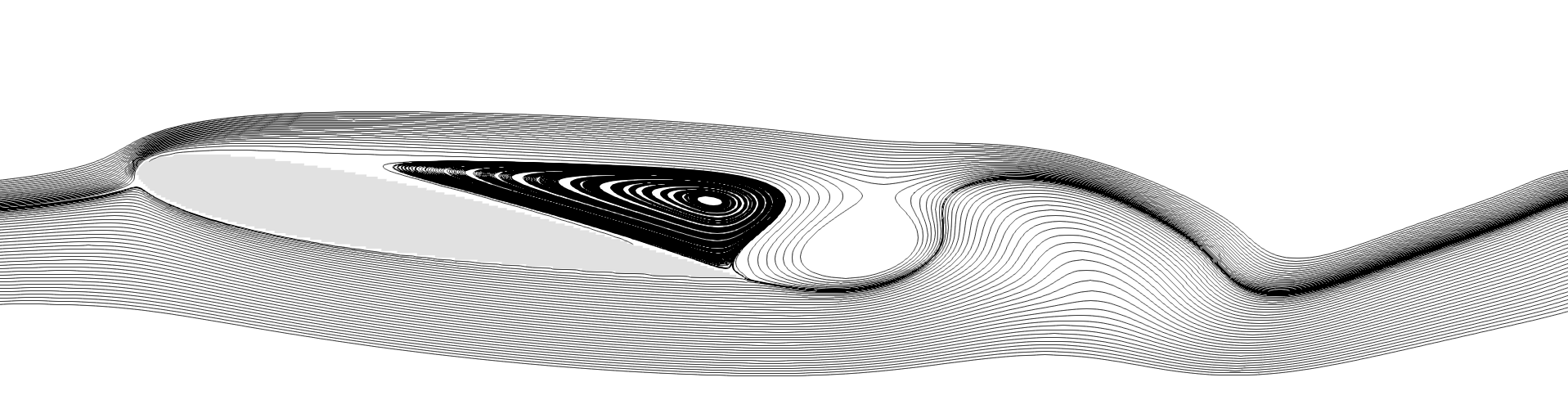}}
\subfigure[\label{naca24} $\alpha=24^{\circ}$]{\includegraphics[width=6.5cm]{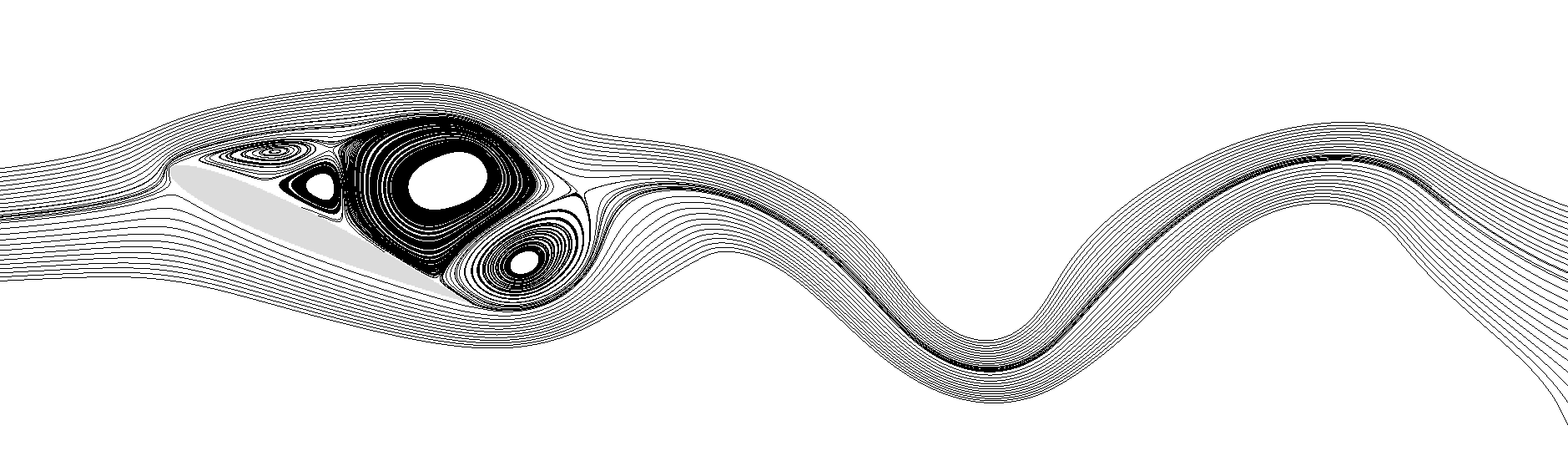}}
\subfigure[\label{naca26} $\alpha=26^{\circ}$]{\includegraphics[width=6.5cm]{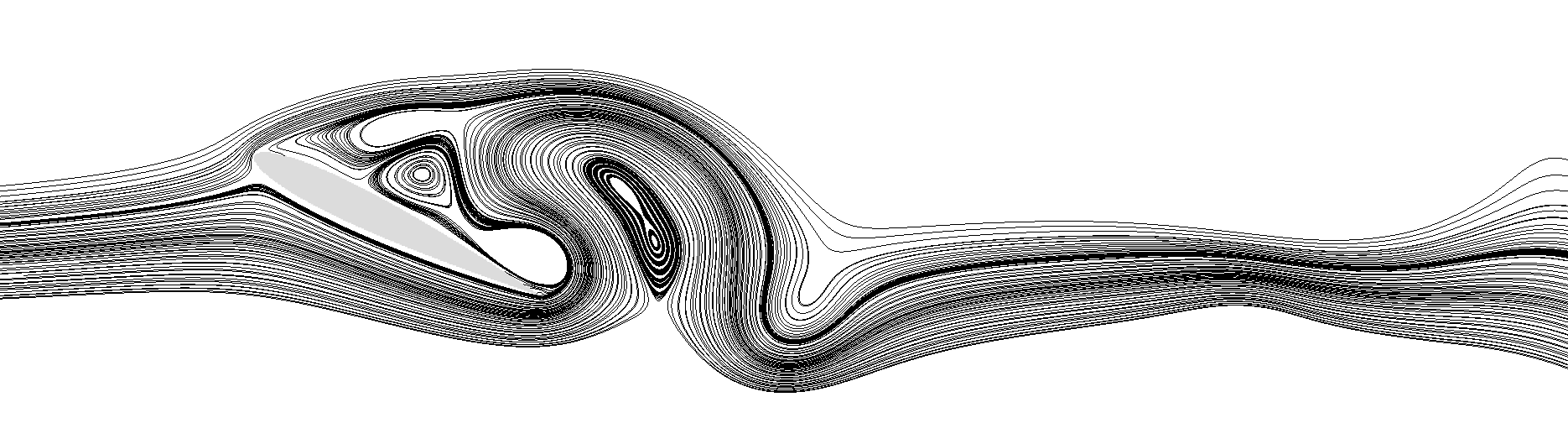}}
\caption{\label{NACAangles} Streamlines around NACA 0012 for various angles }
\end{figure}

For $\alpha \leq 8^{\circ}$ a steady solution has been obtained, which leads to a small increase in drag. After, up to $24^{\circ}$, a periodic vortex shedding  is observed (figures (\ref{naca10}) and (\ref{naca24}). During this phase, the increase in drag is constant, and the lift increase remains fairly stable. An irregularity in the lift coefficient appears at $26^{\circ}$. This stall phenomena is well captured with the VP-LBM approach, and the numerical value is also well computed with ME as well SI. 

Note that the values obtained here are a little higher than those obtained by Kurtulus \cite{Kurtulus2015} and Di Illio \textit{et al.} \cite{DIILIO2018200}, but smaller than those obtained by Liu \textit{et al.}\cite{LIU20123427}. Due to the enclosure of our results with those of Liu  \textit{et al.} \cite{LIU20123427} and Di Illio \cite{DIILIO2018200}, they can be considered as validated. Considering that the work of Di Illio \textit{et al.} \cite{DIILIO2018200} is the reference one, because a finer mesh is used, the Stress Integration method gives better results than the Momentum Exchange method. SI allows a better approximation of the surface, with a direct discretization of the solid boundary, and a true outward normal, while MEM used a staircase approximation. The limit of SI, which is an extrapolation of the distribution values on the boundary, seems to have no consequences in these cases. 

The lift forces are quite similar, ME or SI does not affect the results. The drag has been slightly overestimated with ME, probably because of the approximation of the computational boundary induced by this method.

\subsection{Sedimentation of a particle under gravity}

\begin{figure}[!htpb]
  \centering
  \begin{tikzpicture}
    \draw [line width=5 pt] (0,0) -- (0,4);
 \draw [line width=5 pt] (3.8,0) -- (3.8,4);
 \draw [thick] (1.5,3) circle [radius=0.5];
\draw [right,latex-latex] (4,0) --(4,4);
\node [right] at (4,2) {L};
\draw [below,latex-latex] (0,-0.1) -- (3.8,-0.1);
\node [below] at (2,0) {$H$};
\draw [latex-latex] (1.1,2.7) -- (1.9,3.27);
\draw (1.9,3.27) -- (2.5,3.28);
\node [above] at (2.5,3.28) {D};
\draw [latex-latex] (-1.5,0.5)--(-1.5,1.5) -- (-0.5,1.5);
\node [above] at (-0.5,1.5) {$\mathbf{x}$};
\node [right] at (-1.5,0.5) {$\mathbf{y}$};
\draw [latex-latex] (0,2.4) -- (1.5,2.4);
\node [below] at (0.75,2.5) {$x_0$};
\draw [latex-latex] (0.9,3) -- (0.9,4);
\node [left] at (0.9,3.5) {$y_0$};

\draw [red,line width=2 pt,-latex] (3,3) -- (3,2);
\node [right] at (3,2.5) {$\boldsymbol{g}$};
\end{tikzpicture}
\caption{Schematic description of particle sedimentation}
  \label{schemsuspension}
\end{figure}
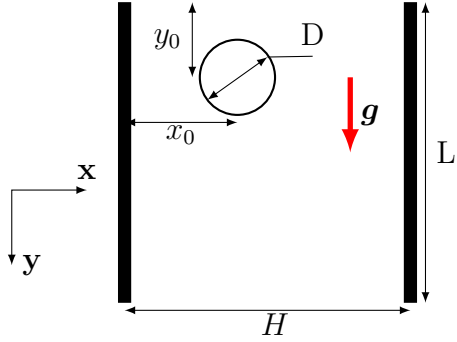

The next case focuses to the sedimentation of a particle under gravity in an infinite channel (figure \ref{schemsuspension}) for non-centered configurations. This problem has been widely used for model validations and is very useful for testing the ability of a method to capture complex trajectories \cite{Wang20131151,Tao20161,WEN2014161,Li2004}.

A circular particle of diameter $D$ falls by gravity $\boldsymbol{g}$ into a fluid of density $\rho$ in a vertical channel of width $H$. In the initial state, the particle is at a distance $x_0$ from the left wall, a distance  $y_0$ from the top of the channel and the velocity of the particle is equal to zero. In this case, the displacement of the particle can be described using the equations (\ref{dep1}) and (\ref{theta2}) :
\begin{eqnarray}
m \dfrac{d^2 \mathbf{x_G}}{d t^2} &=&\boldsymbol{\mathcal{F}_f} + m\left( 1 - \dfrac{\rho}{\rho_s} \right) \boldsymbol{g} \label{dep1}\\
I \dfrac{d^2 \boldsymbol{\theta}}{d t^2}  &=& \boldsymbol{\mathcal{T}_f} \label{theta2}
\end{eqnarray}
where $\rho$ denotes the fluid density, $\rho_s$ the solid density and $m$ the particle mass. The last term of the equation (\ref{dep1}) represents the weight and buoyancy (Archimedes' principle) acting on the particle.

For small Reynolds numbers and a large non dimensional width $\tilde{H}=\dfrac{H}{D}$, the particle reaches a steady state.

The following case deals with a particle whose initial position is not at the center of the channel ($x_0=0.75 D$). The properties of the fluid are  $\rho=1 \;\textrm{g}\cdot\textrm{cm}^{-3}$, and $\mu=0.1 \; \textrm{g}\cdot \textrm{cm}^{-1} \cdot\textrm{s}^{-1}$ and the physical problem concerns a particle of diameter $D=0.1 \; \textrm{cm}$. Four mass ratio $\rho_r =\dfrac{\rho_s}{\rho_f}= 1.0015,1.003, 1.0015$ and $1.03$ and $\Vert  \boldsymbol{g} \Vert = 980 \; \textrm{cm} \cdot \textrm{s}^{-2}$ are used. The Reynolds numbers based on the final velocity of the particle are, respectively, $\textrm{Re} = 0.52, 1.03, 3.23$ and $8.33$.

For the LBM computations the cylinder diameter was $26 \; \textrm{l.u.}$, the same value used in the literature \cite{Tao20161}, the relaxation time was $\tau=0.6$. No-slip boundary conditions were imposed on the left and right walls. A zero velocity boundary condition was applied at the inlet (top of the channel) and free flow conditions were applied at the outlet (bottom). A large value of $L$ was chosen, so that the inlet and the outlet do not influence the behavior of the particle.



\begin{figure}[!htbp]
\center
\includegraphics[width=10cm]{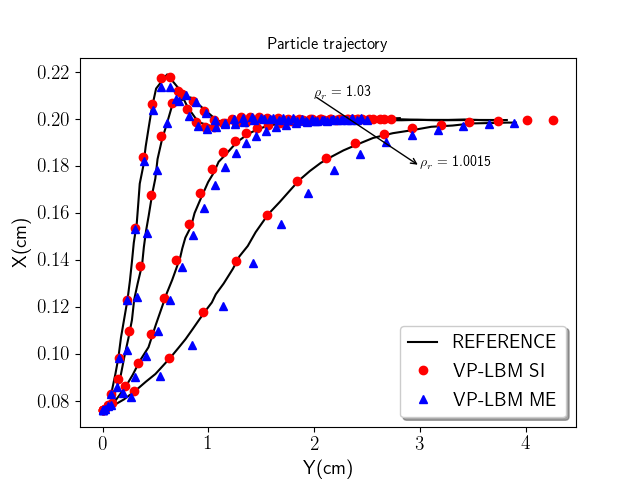}
\caption{\label{pt} Results obtained using the VP-LBM approach and compared with Tao et al's results \cite{Tao20161}}
\end{figure}

The particle trajectory for each mass ratio is plotted in figure \ref{pt}, and compared with the reference results of the literature \cite{Tao20161,Li2004}. In order to facilitate the reading of the figure, only a few points for each trajectory have been plotted. First of all, it can be noted that the VP-LBM method, coupled with the Stress Integration method gives for each case a good behavior of the particle. The results are similar to those obtained with the UIBB and the literature. This is not what is observed for VP-LBM coupled with Momentum Exchange. The trajectory is almost correct for a high mass ratio, although a small difference can be observed around $t=0.5 \textrm{s}$ for $\rho_r=1.03$ , and the error increases as the mass ratio (i.e the Reynolds number) decreases. Our analysis is that for small mass ratio, the fluid forces are very small, and a small error has a greater significance in the behavior of the particle than for a larger mass ratio. The lack of accuracy of the fluid solid interface has a great consequence here. 

\begin{figure}[!htbp]
\center 
\subfigure[\label{rv1} $\rho_r=1.0015$]{\includegraphics[width=6cm]{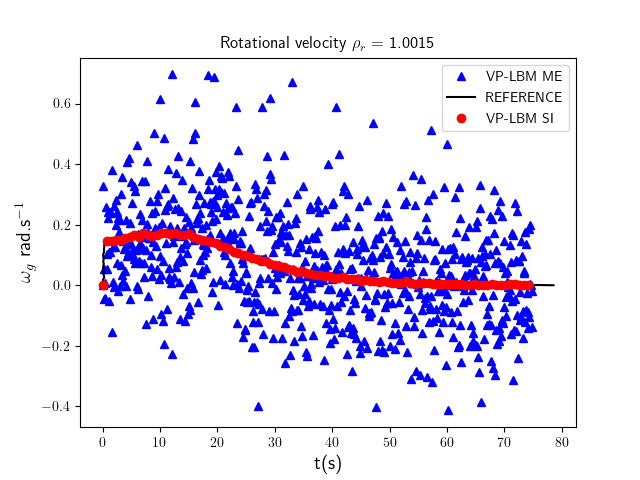}}
\subfigure[\label{rv4} $\rho_r=1.03$]{\includegraphics[width=6cm]{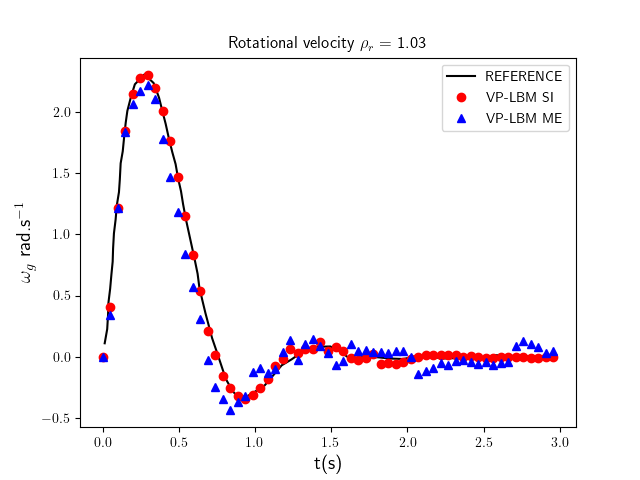}}
\caption{\label{rv} Rotational velocity obtained using the VP-LBM approach and compared with reference's results \cite{Tao20161,Li2004}}
\end{figure}

Figures \ref{rv} plot the rotational velocity for the smallest and largest mass ratio. For $\rho_r=1.0015$ (figure \ref{rv1}) spurious oscillations are observed with ME. Even if the average follows the reference solutions, these oscillations lead to particle deviation from the reference trajectory. In the figure \ref{rv4}, oscillations are smaller, but even if the solution is close to the reference one, the Stress Integration gives better results.

\begin{figure}[!htbp]
\center
\subfigure[\label{u1} t = 0.4 s]{\includegraphics[width=3cm]{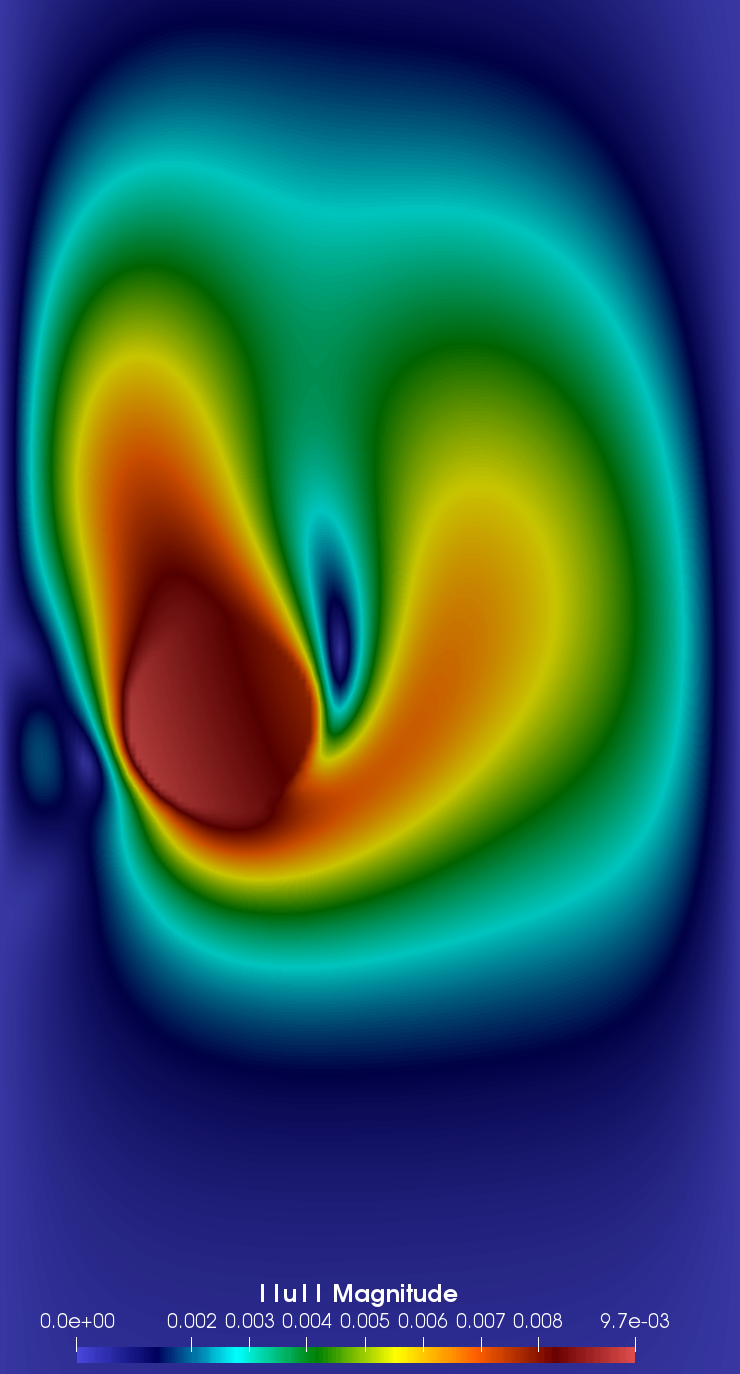}}
\subfigure[\label{u2} t = 0.6 s]{\includegraphics[width=3cm]{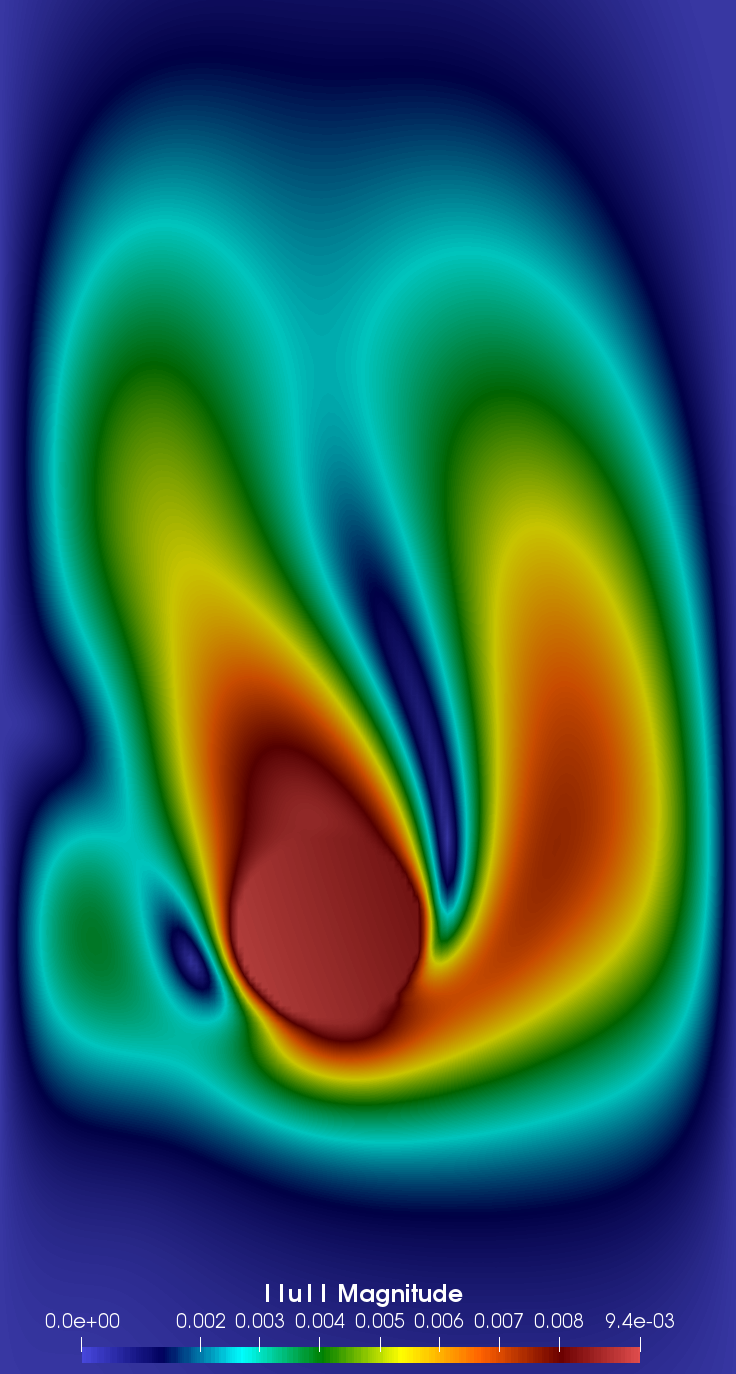}}
\subfigure[\label{u3} t = 1.0 s]{\includegraphics[width=3cm]{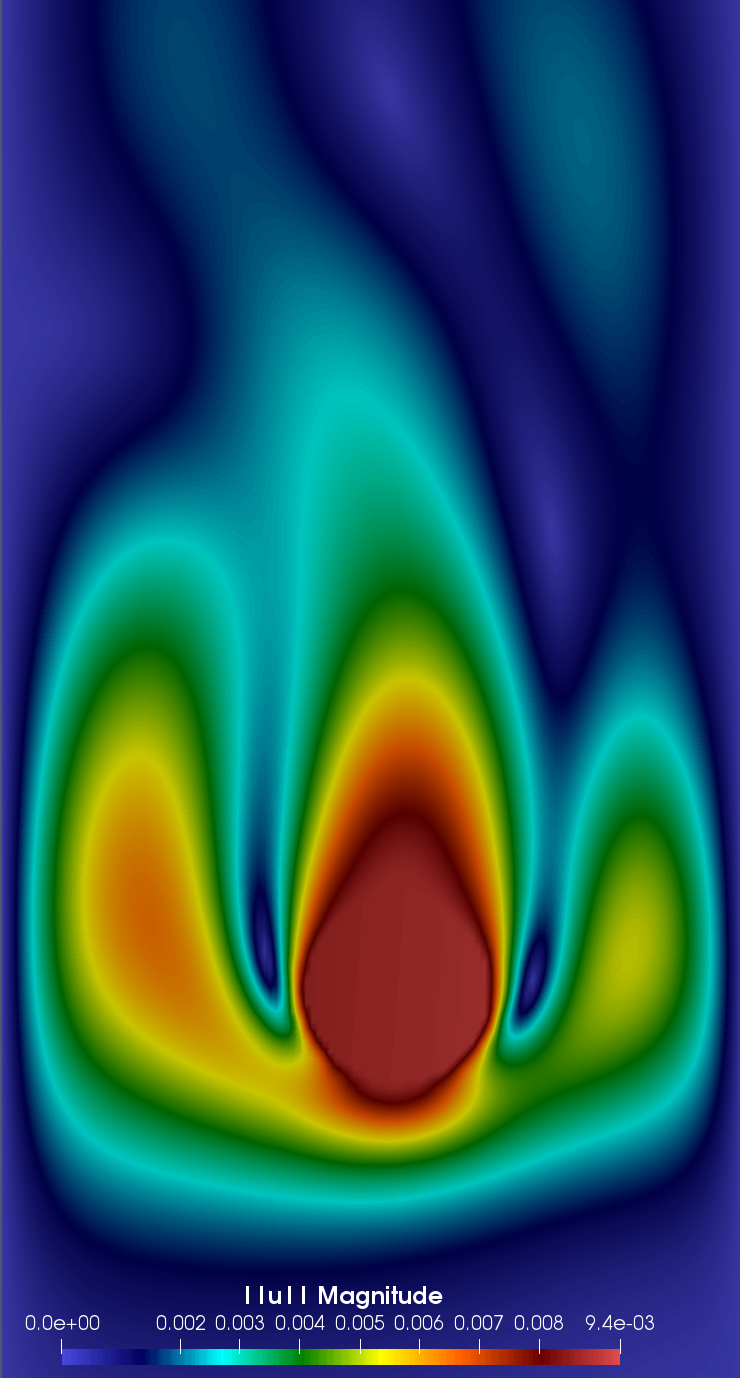}}
\subfigure[\label{u4} t = 3.0 s]{\includegraphics[width=3cm]{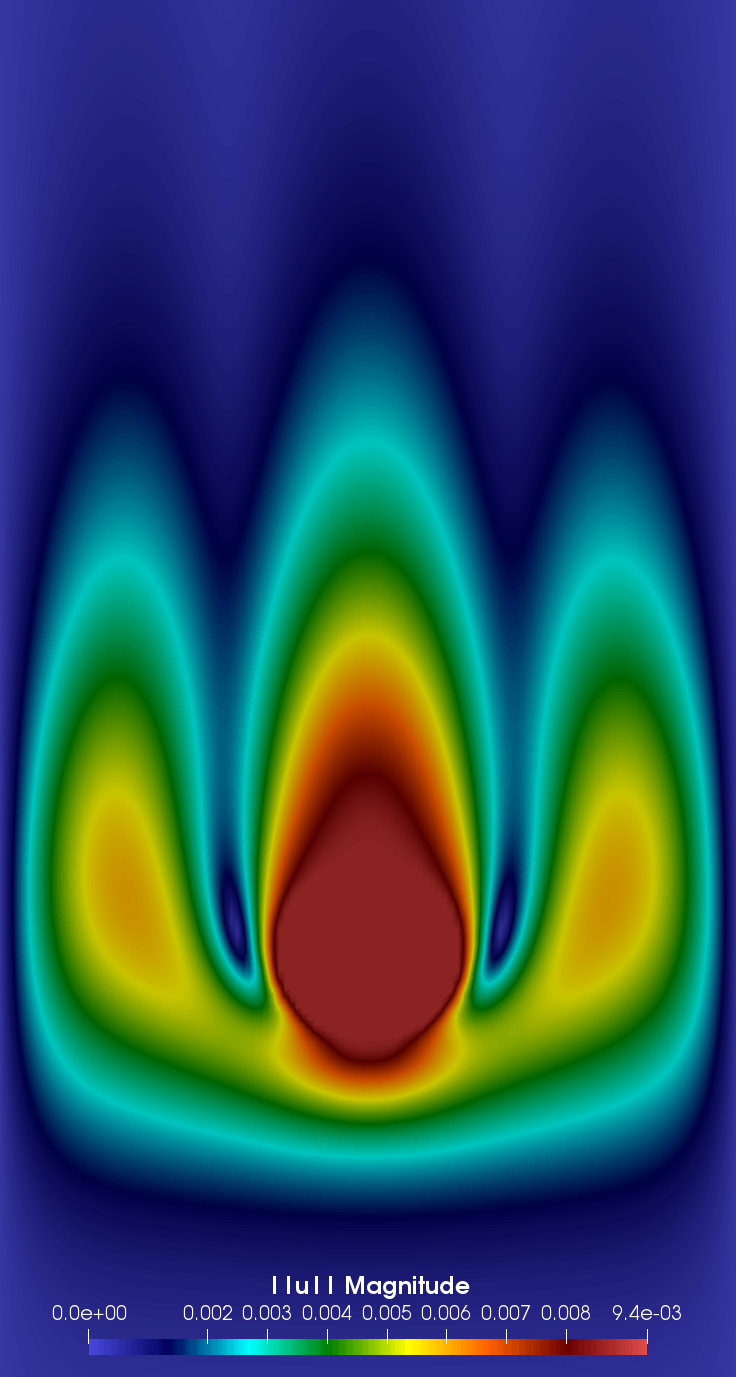}}
\caption{\label{vnc} Fluid velocity magnitude at times t=0.4, 06, 1.0 and 3.0 seconds in lattice units}
\end{figure}

\begin{figure}[!htbp]
\center
\subfigure[\label{v1} t = 0.4 s]{\includegraphics[width=3cm]{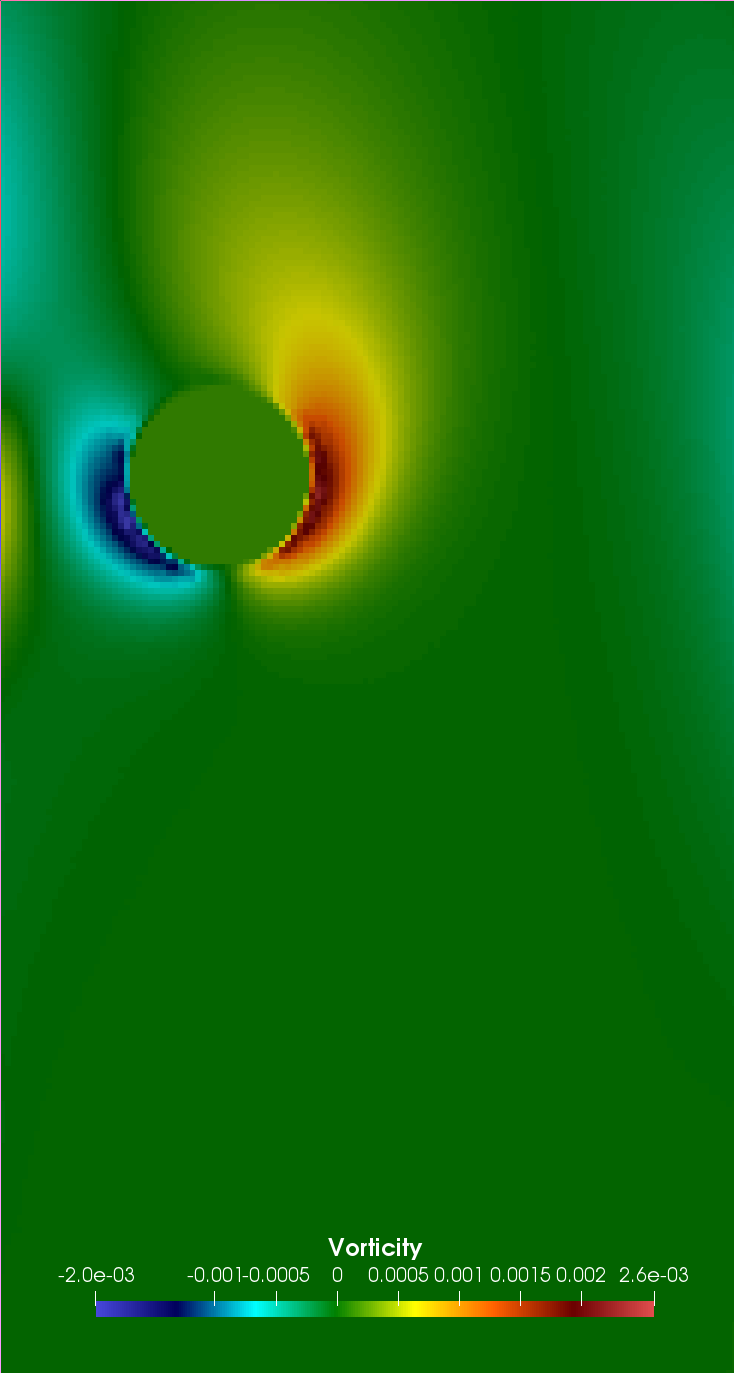}}
\subfigure[\label{v2} t = 0.6 s]{\includegraphics[width=3cm]{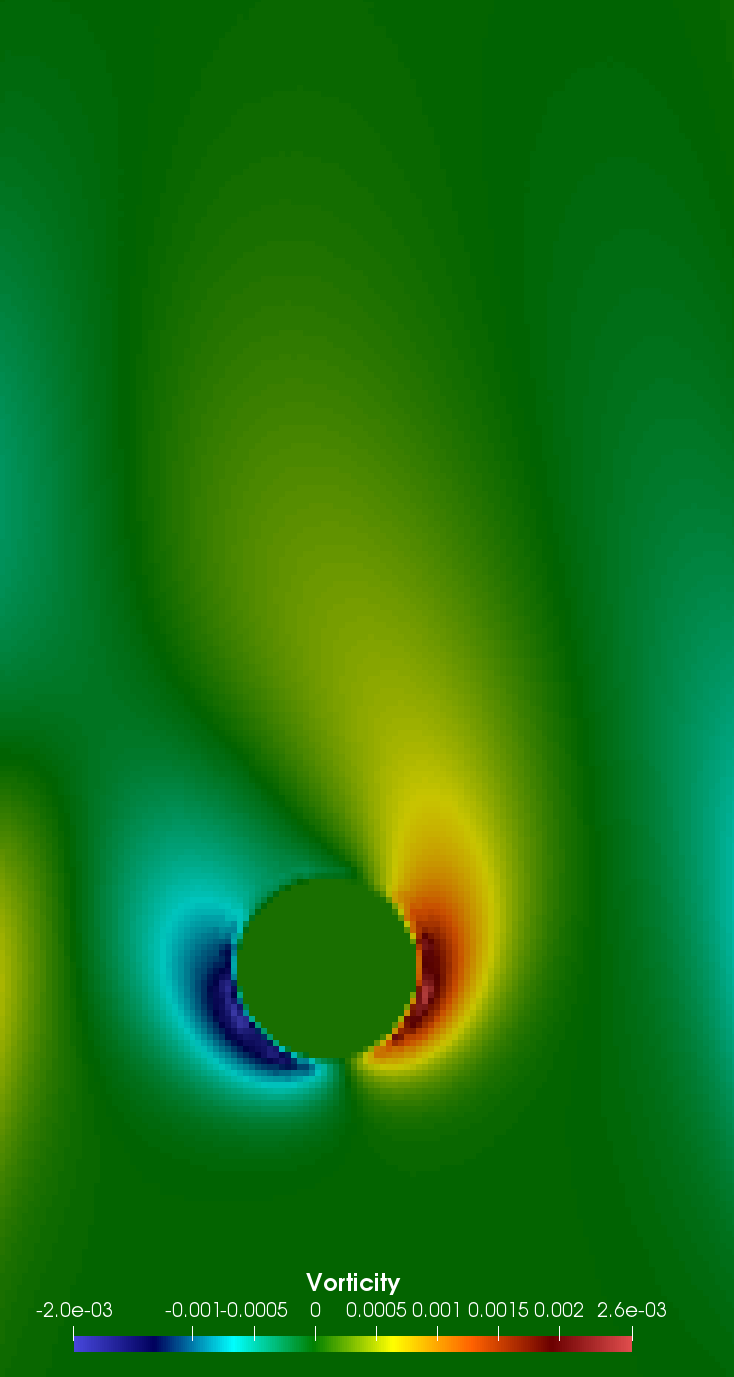}}
\subfigure[\label{v3} t = 1.0 s]{\includegraphics[width=3cm]{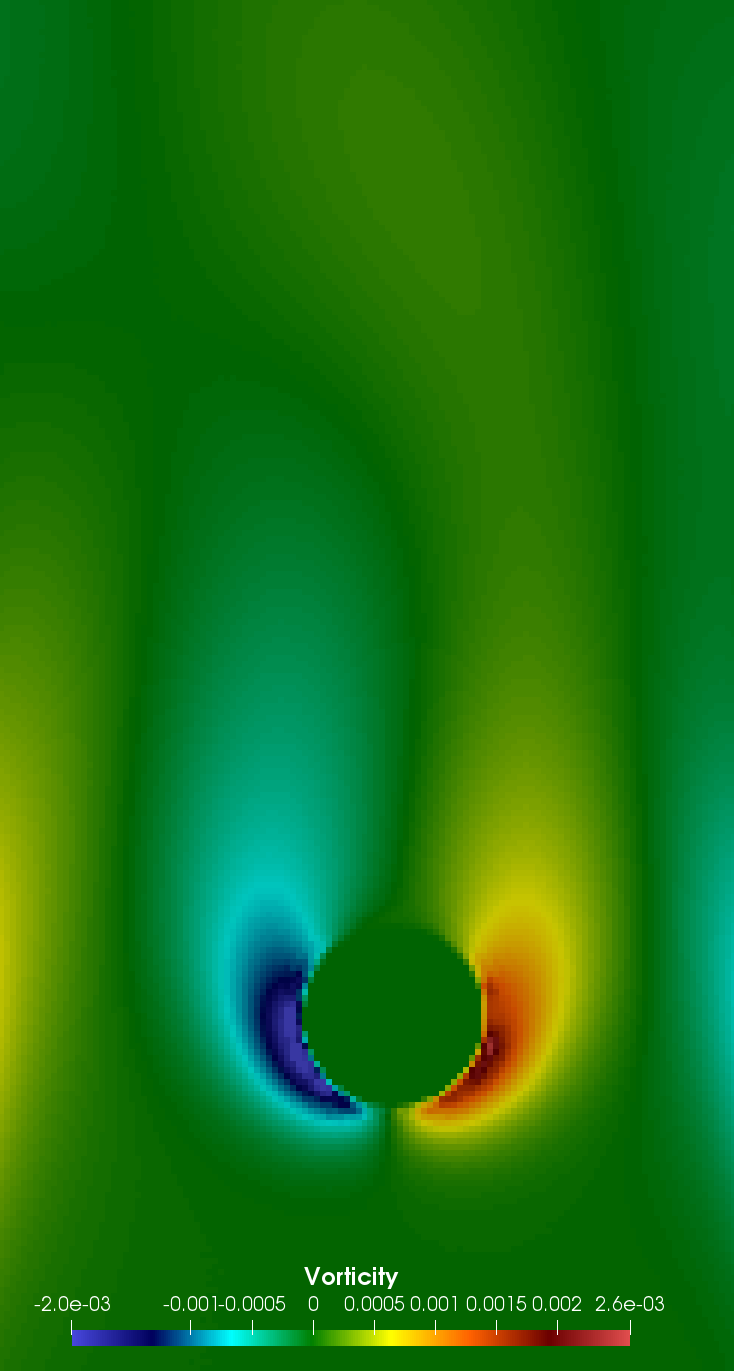}}
\subfigure[\label{v4} t = 3.0 s]{\includegraphics[width=3cm]{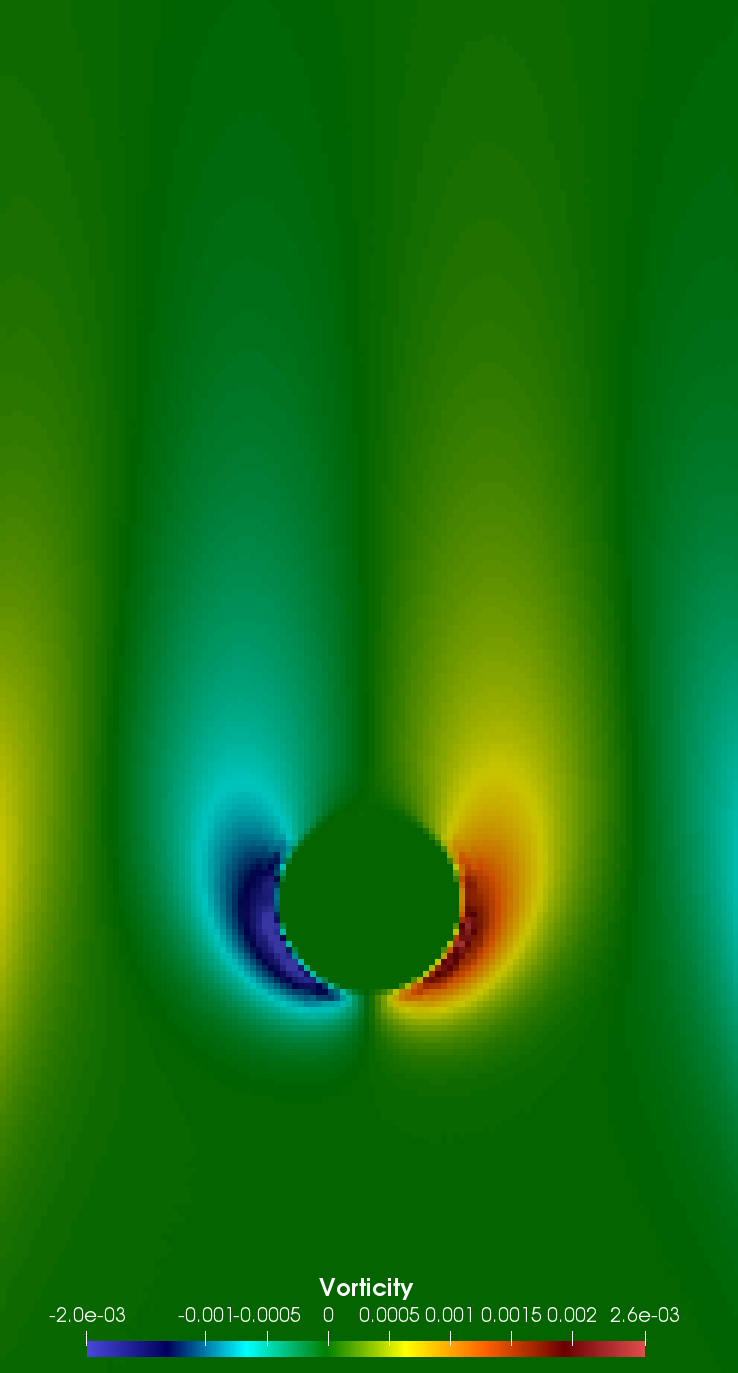}}
\caption{\label{vortnc} Fluid vorticity  at times t=0.4, 06, 1.0 and 3.0 seconds in lattice units}
\end{figure}

The figures \ref{vnc} and \ref{vortnc} show the fluid velocity and the vorticity field around the particle at four different times. The dynamics of the flow field and the particle can be analyzed using the velocity magnitude and the vorticity. The particle goes first to the right and rotates in a positive direction. Next a brief oscillation occurs around the central line of the channel and finally the particle stays in the middle of the channel with a steady velocity.

This example shows that the VP-LBM method is able to predict a complex trajectory for a real case of fluid structure interaction at a very low Reynolds number.

\section{Conclusion}
The Volume Penalization method coupled with Lattice Boltzmann method (VP-LBM) was successfully applied to two new cases. The methods available for fluid loads  calculation have been discussed. The VP-LBM has shown its ability to reproduce the complex physics of an airfoil at different angles of attack, and the stall phenomenon has been well captured. For this application, the Momentum Exchange (ME) and the Stress Integration (SI) methods give similar results, but the drag coefficients seem a little bit more accurate with SI. In the second example, the particle sedimentation under gravity, the SI method has given the best results. The trajectories have been perfectly recovered, spurious oscillations observed with the ME method have been cancelled with SI. VP-LBM combined with the stress integration method seems to be a valid tool to simulate fluid structure interaction problems.



\section{Bibliography}
\bibliographystyle{elsarticle-num} 
\bibliography{biblbmvp}






\end{document}